\documentstyle[aps,eqsecnum]{revtex}

\newcommand{\be}{\begin{eqnarray}}
\newcommand{\ee}{\end{eqnarray}}
\newcommand{\la}{\langle}
\newcommand{\ra}{\rangle}
\newcommand{\bfx}{{\bf x}}

\newcommand{\bfp}{{\bf p}}
\newcommand{\veps}{\varepsilon}

\begin{document}
\title{ Relativistic recoil corrections to the atomic energy levels
}

\author{V. M. Shabaev \\  Department of Physics, St. Petersburg    
State University,\\
 Oulianovskaya Street 1, Petrodvorets,    
St. Petersburg 198504, \\ Russia}

\maketitle              % typesets the title of the contribution

\begin{abstract}

The quantum electrodynamic theory of the nuclear recoil
effect in atoms to all orders in $\alpha Z$ and to first
order in $m/M$ is considered. The complete $\alpha Z$-dependence
formulas for the relativistic recoil corrections to the atomic 
energy levels are derived in a simple way.  The results of numerical 
calculations of the recoil effect to all orders in $\alpha Z$ are 
presented for hydrogenlike and lithiumlike atoms. These results 
are compared with analytical results obtained to lowest orders 
in $\alpha Z$. It is shown that even for hydrogen the numerical 
calculations to all orders in $\alpha Z$ provide most precise
theoretical  predictions for the relativistic recoil correction
of first order in $m/M$.

\end{abstract}

%\tableofcontents    

%\newpage    
%%%%%%%%%%%%%%%%%%%%%%%%%%%   
\section{Introduction}    
%%%%%%%%%%%%%%%%%%%%%%%%%%%

In the non-relativistic quantum mechanics the nuclear recoil
effect for a hydrogenlike atom is easily taken into account
by using the reduced mass $\mu=mM/(m+M)$ instead of the electron
mass $m$ ($M$ is the nuclear mass). 
It means that to account for the nuclear
recoil effect to first order in $m/M$ we must simply replace 
the binding energy $E$ by $E(1-m/M)$. 

Let us consider now a relativistic hydrogenlike atom.
 In the infinite nucleus mass approximation
a hydrogenlike atom is described by the Dirac equation
($\hbar=c=1$)
   \begin{eqnarray}
 (-i\mbox{\boldmath $\alpha$}\cdot \mbox{\boldmath $\nabla$}
+\beta m+V_{C}(\bfx))\psi(\bfx)= 
 \varepsilon \psi(\bfx)\,,
 \label{dirac}
    \end{eqnarray}  
where $V_{C}$ is the Coulomb potential of the nucleus.
For the point-nucleus case, analytical solution of this
equation yields the well known formula for the energy
of a bound state:
\begin{eqnarray}\label{diren}
\veps_{nj}=\frac{mc^2}{\sqrt{1+\frac{(\alpha Z)^2}
{[n-(j+1/2)+\sqrt{(j+1/2)^2-(\alpha Z)^2}\;]^2}}}\,,
\end{eqnarray}
where $n$ is the principal quantum number and
$j$ is the total angular momentum of the electron.
The main problem we will discuss in this paper
is the following: what is the recoil correction to this
formula?

It is known that to the lowest order in $\alpha Z$ the 
relativistic recoil
correction to the energy levels can be derived from the
Breit equation. Such a derivation was made by Breit and Brown in
1948 \cite{breit48} (see also \cite{bechert35}).
 They found that the relativistic recoil
correction to the lowest order in $\alpha Z$ consists of two terms.
The first term reduces the fine structure splitting by the
factor $(1-m/M)$. The second term does not affect the fine structure
splitting and is equal to $-(\alpha Z)^4m^2/(8Mn^4)$.
Calculations of the recoil effect to higher orders in $\alpha Z$
demand using QED beyond the Breit approximation. In quantum
electrodynamics a two-body system is generally treated by the
Bethe-Salpeter method \cite{salpeter51} or by one of versions
of the quasipotential method proposed first by Logunov and
Tavkhelidze \cite{logunov63}. In Ref. \cite{salpeter52} (see also
\cite{bethe57}), using the Bethe-Salpeter
equation, Salpeter calculated the recoil correction
of order $(\alpha Z)^5 m^2/M$ to the
energy levels of a hydrogenlike atom. 
This correction gives a contribution of 359 kHz to the 
$2s$ - $2p_{1/2}$ splitting in hydrogen. The current 
uncertainties of the Lamb and isotopic shift measurements 
are much smaller than this value (see, e.g.,
\cite{hansch00}) and, therefore, calculations of the
recoil corrections of higher orders in $\alpha Z$
are required. In addition, for the last decade a great progress was made
in high precision measurements of the Lamb shifts in high-Z few-electron
ions \cite{schweppe91,beiersdorfer98,stoehlker00}. In these systems, the
parameter $\alpha Z$ is not small and, therefore, calculations of 
the relativistic recoil corrections
to all orders in $\alpha Z$ are needed.

%%%%%%%%%%%%%%%%%%%%%%%%%%%%%    
\section{Relativistic formula for the recoil correction}    
%%%%%%%%%%%%%%%%%%%%%%%%%%%%%   

First attempts to derive formulas for 
the relativistic recoil corrections
to all orders in $\alpha Z$ were undertaken in
\cite{labzowsky72,braun73}. As a result
of these attempts, only a part of the desired expressions
 was found in \cite{braun73} (see Ref. \cite{shabaev98} for details).
The complete $\alpha Z$-dependence formula for the relativistic
recoil effect in the case of a hydrogenlike atom was 
derived in \cite{shabaev85}.
The derivation of \cite{shabaev85} was based
on using a quasipotential equation
in which the heavy particle is put on the mass shell
\cite{gross69,faustov75}.
According to \cite{shabaev85}, the relativistic recoil
correction to the energy of a state $a$
is the sum of a 
lower-order term $\Delta E_{\rm L}$
and a higher-order term $\Delta E_{\rm H}$:
\begin{eqnarray} \label{recsh}
\Delta E &=&\Delta E_{\rm L}+\Delta E_{\rm H}\,,\\
\Delta E_{\rm L}&=&\frac{1}{2M}\langle a|\Bigl(
{\bf p}^2-
\frac{\alpha Z}{r}\Bigl(
\mbox{\boldmath $\alpha$}+\frac
{(\mbox{\boldmath $\alpha$}
\cdot{\bf r}){\bf r}}{r^{2}}
\Bigr)\cdot{\bf p}\Bigr)|a\rangle\,,
\label{reclo}
\\
\Delta E_{\rm H}&=&\frac{i}{2\pi M}\int_{-\infty}^{\infty}d\omega\,
\langle a|\Bigl({\bf D}(\omega)-
\frac{[{\bf p},V_{\rm C}]}{\omega+i0}
\Bigr)\nonumber\\
&&\times
G(\omega +\varepsilon_{a})
\Bigl({\bf D}(\omega)+\frac{[{\bf p},V_{\rm C}]}{\omega+i0}\Bigr)
|a\rangle\,. \label{recho}
\end{eqnarray}
Here $|a\rangle$ is the unperturbed state of the Dirac electron
in the Coulomb field $V_{\rm C}(r)=-\alpha Z/r$,
${\bf p}=-i\mbox{\boldmath $\nabla$} $ 
is the momentum operator, 
 $G(\omega)=[\omega-H(1-i0)]^{-1}$ is the relativistic
Coulomb-Green function, 
$ H=\mbox{\boldmath $\alpha$}\cdot
{\bf p}+\beta m +V_{\rm C}\,$, 
$\alpha_{l}\;(l=1,2,3)$ are the Dirac matrices,
$\varepsilon_{a}$ is the unperturbed Dirac-Coulomb energy,
\be \label{ddd}
D_{m}(\omega)=-4\pi\alpha Z\alpha_{l}D_{lm}(\omega)\,,
\ee
$ D_{lm}(\omega)$ is
the transverse part of the photon propagator in the Coulomb gauge.
In the coordinate representation it is
\be
D_{ik}(\omega,{\bf r})=-\frac{1}{4\pi}\Bigl\{\frac
{\exp{(i|\omega|r)}}{r}\delta_{ik}+\nabla_{i}\nabla_{k}
\frac{(\exp{(i|\omega|r)}
-1)}{\omega^{2}r}\Bigr\}\,.
\ee
The scalar product is implicit in the equation (\ref{recho}). 
In Refs. \cite{yelkhovsky94,pachucki95}, the formulas 
(\ref{recsh})-(\ref{recho}) were rederived 
by other methods and in \cite{yelkhovsky94} it was noticed that 
$\Delta E$ can be written  in the following compact form:
\be \label{yel}
\Delta E=\frac{i}{2\pi M}\int_{-\infty}^{\infty}d\omega\,
\langle a|[\bfp- {\bf D}(\omega)]
G(\omega +\varepsilon_{a})
[\bfp-{\bf D}(\omega)]|a\rangle\,. \label{yelkh}
\end{eqnarray}
However, the representation (\ref{recsh})-(\ref{recho}) is more
convenient for practical calculations.

The term $\Delta E_{\rm L}$ can easily be calculated by using the virial
relations for the Dirac equation \cite{epstein62,shabaev91}.
 Such a calculation  gives \cite{shabaev85}
\begin{eqnarray} \label{shlo}
\Delta E_{\rm L}=\frac{m^{2}-\varepsilon_{a}^{2}}{2M}\,.
\end{eqnarray}
This simple formula contains all the recoil corrections
within the $(\alpha Z)^{4}m^2/M$ approximation.
The term $\Delta E_{\rm H}$ taken to the
lowest order in $\alpha Z$ gives the Salpeter  correction \cite{salpeter52}.
Evaluation of this term to all orders in $\alpha Z$  will be discussed
below.

The complete $\alpha Z$-dependence formulas for the nuclear recoil
corrections in high $Z$ few-electron atoms were derived 
in Ref. \cite{shabaev88}.
As it follows from these formulas, within
 the $(\alpha Z)^{4}m^{2}/M$
approximation the nuclear  recoil corrections
 can be obtained
by averaging the operator 
\begin{eqnarray} \label{hlo}
H_{M}^{(\rm L)}
=\frac{1}{2M}\sum_{s,s'}\Bigl({\bf p}_{s}\cdot{\bf p}_{s'}-
\frac{\alpha Z}{r_{s}}\Bigl(\mbox{\boldmath $\alpha$}_{s}+\frac
{(\mbox{\boldmath $\alpha$}_{s}\cdot{\bf r}_{s}){\bf r}_{s}}{r_{s}^{2}}
\Bigr)\cdot{\bf p}_{s'}\Bigr)
\end{eqnarray}
with the Dirac wave functions.
This operator can also be used for relativistic
calculations of the nuclear recoil
effect in neutral atoms.
An independent derivation of this operator was done in \cite{palmer87}.
The operator (\ref{hlo}) was employed in \cite{shabaev94} to calculate
the $(\alpha Z)^{4}m^{2}/M$ corrections
to the energy levels of two-
 and three-electron multicharged ions.

%%%%%%%%%%%%%%%%%%%%%%%%%%%%%%%%%%%%%%%%%%%%
\section{Simple approach to the recoil effect in atoms}
%%%%%%%%%%%%%%%%%%%%%%%%%%%%%%%%%%%%%%%%%%%

 As was shown
in \cite{shabaev98}, to include the relativistic
recoil corrections in calculations of the
energy levels, we must add to the standard Hamiltonian
of the electron-positron field interacting with the quantized
electromagnetic field and with the Coulomb field of the nucleus
$V_{\rm C}$, taken in the Coulomb gauge, the following term
\begin{eqnarray} \label{hamm}
H_{M}&=&
\frac{1}{2M}
\int d{\bf x}
\psi^{\dag}({\bf x})(-i
\nabla_{\bf x})
\psi({\bf x})
\int d{\bf y}
\psi^{\dag}({\bf y})(-i
\nabla_{\bf y})
\psi({\bf y})\nonumber\\
&&-\frac{eZ}{M}
\int d{\bf x}
\psi^{\dag}({\bf x})(-i
\nabla_{\bf x})
\psi({\bf x}){\bf A}(0)+\frac{e^{2}Z^{2}}{2M}{\bf A}^{2}(0)\,.
\end{eqnarray}
This operator acts only on the electron-positron and 
electromagnetic field variables.
The normal ordered form of $H_M$ taken in the interaction
representation must be added to the interaction Hamiltonian.
It gives additional elements to the Feynman rules for the Green 
function. In the Furry picture, in addition
to the standard Feynman rules  in the energy representation
(see \cite{shabaev94a,shabaev98}),
the  following verteces and lines appear (we assume that 
the Coulomb gauge is used)

\begin{enumerate}
\item {\it Coulomb contribution}.\\

An additional line ("Coulomb-recoil" line) appears to be
\newline\\
\setlength{\unitlength}{0.7mm}
\begin{picture}(60,5)(0,0)
  \multiput(15,2)(2,0){15}{\circle*{1} }
  \put(15,2){\circle*{2}}
  \put(45,2){\circle*{2}}
  \put(30,6){$\omega$}
  \put(12,-6){${\bf x}$}
  \put(45,-6){${\bf y}$}
  \label{intphotline}
\end{picture}
          $ \frac{i}{2\pi} \frac{\delta_{kl}}{M}
\int_{-\infty}^{\infty}d\omega\,. $ \\
\newline
This line joins two vertices each of which corresponds to
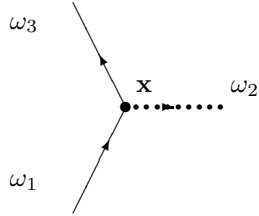
\newline\\
\begin{picture}(60,60)(0,0)
  \put(50,30){\line(-1,2){10}}
  \put(50,30){\line(-1,-2){10}}
  \multiput(50,30)(2,0){10}{\circle*{1}}
  \put(50,30){\circle*{2}}
  \put(52,33){{\bf x}}
  \put(70,33){$\omega_{2}$}
  \put(28,15){$\omega_{1}$}
  \put(28,45){$\omega_{3}$}
  \put(58,30){\vector(1,0){1}}
  \put(46,38){\vector(-1,2){1}}
  \put(46,22){\vector(1,2){1}}
\label{vertex}
\end{picture}  
       $ - 2\pi i\gamma^{0}\delta(\omega_{1}-\omega_{2}-\omega_{3})
            \int d{\bf x}\,p_{k} \;,$ \newline \\ 
\newline
where
 ${\bf p}=
-i\nabla_{\bf x}$ and $k=1,2,3$.

\item {\it One-transverse-photon contribution}.\\

An additional vertex on an electron line appears to be
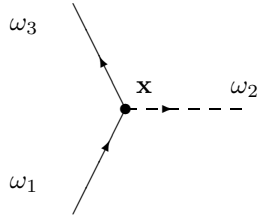
\newline
\begin{picture}(60,60)(0,0)
  \put(50,30){\line(-1,2){10}}
  \put(50,30){\line(-1,-2){10}}
  \multiput(50,30)(4,0){6}{\line(1,0){2}}
  \put(50,30){\circle*{2}}
  \put(52,33){{\bf x}}
  \put(70,33){$\omega_{2}$}
  \put(28,15){$\omega_{1}$}
  \put(28,45){$\omega_{3}$}
  \put(58,30){\vector(1,0){1}}
  \put(46,38){\vector(-1,2){1}}
  \put(46,22){\vector(1,2){1}}
\label{vertex1}
\end{picture}  
   $ - 2\pi i\gamma^{0}
\delta(\omega_{1}-\omega_{2}-\omega_{3})
       \frac{eZ}{M} \int d{\bf x}\,p_{k} \;,$ \newline \\ 

The transverse photon line attached to this vertex (at
the point ${\bf x}$) is
\newline\\
\begin{picture}(60,5)(0,0)
  \multiput(15,2)(4,0){9}{\line(1,0){2} }
  \put(15,2){\circle*{2}}
  \put(30,6){$\omega$}
  \put(12,-6){${\bf x}$}
  \put(47,-6){${\bf y}$}
  \label{intphotline1}
\end{picture}
          $ \frac{i}{2\pi}
\int_{-\infty}^{\infty}d\omega D_{kl}(\omega,{\bf y})\,. $ \\
\newline\\
At the point ${\bf y}$ this line is to be attached to an usual
vertex in which we have $-2\pi i e\gamma^{0}\alpha_{l}2\pi
\delta(\omega_{1}-\omega_{2}-\omega_{3})\int d{\bf y}$,
where $\alpha_{l}$ ($l=1,2,3$)
are the usual Dirac matrices.

\item {\it Two-transverse-photon contribution}.\\

An additional line ("two-transverse-photon-recoil" line)
appears to be
\newline\\
\begin{picture}(60,5)(0,0)
  \multiput(15,2)(4,0){9}{\line(1,0){2} }
  \put(32,2){\circle*{2}}
  \put(12,-6){${\bf x}$}
  \put(47,-6){${\bf y}$}
  \put(30,6){$\omega$}
  \label{intphotline1}
\end{picture}
          $ \frac{i}{2\pi}\frac{e^{2}Z^{2}}{M}
\int_{-\infty}^{\infty}d\omega
 D_{il}(\omega,{\bf x})
 D_{lk}(\omega,{\bf y})\,. $ 
\newline\\
This line joins  usual vertices (see the previous item).
\end{enumerate}

Let as apply this formalism to the case of a single level $a$
in a one-electron atom.
To find the Coulomb nuclear recoil correction we have to calculate 
the contribution of the diagram shown in Fig. 1. 
A simple calculation of this diagram yields (see Ref. \cite{shabaev98}
for details)
\begin{eqnarray}\label{coul}
\Delta E_{\rm C}=\frac{1}{M}\frac{i}{2\pi}
\int_{-\infty}^{\infty}d\omega \sum_{n}\frac
{\langle a| p_{i}|n\rangle\langle n|p_{i}|a\rangle}
{\omega-\varepsilon_{n}(1-i0)}\,.
\end{eqnarray}
The one-transverse-photon nuclear recoil correction corresponds
to the diagrams shown in Fig. 2. 
One easily obtains
\begin{eqnarray} \label{trans1}
\Delta E_{tr(1)}&=&\frac{4\pi\alpha Z}{M}\frac{i}{2\pi}
\int_{-\infty}^{\infty}d\omega\,\sum_{n}\Biggl\{
\frac{
\langle a|p_{i}|n\rangle\langle n|
\alpha_{k} D_{ik}(\varepsilon_{a}-\omega)|a\rangle}
{\omega-\varepsilon_{n}(1-i0)}
\nonumber\\
&&
+\frac{
\langle a|
\alpha_{k} D_{ik}(\varepsilon_{a}-\omega)|n\rangle
\langle n|p_{i}|a\rangle}
{\omega-\varepsilon_{n}(1-i0)}\Biggr\}\,.
\end{eqnarray}
The two-transverse-photon nuclear recoil correction is defined
by the diagram shown in Fig. 3. 
We find
\begin{eqnarray} \label{trans2}
\Delta E_{tr(2)}&=&\frac{(4\pi\alpha Z)^{2}}{M}\frac{i}{2\pi}
\int_{-\infty}^{\infty}d\omega\,\sum_{n}\nonumber\\
&&\times\frac{
\langle a|\alpha_{i}D_{il}(\varepsilon_{a}-\omega)|n\rangle\langle n|
\alpha_{k} D_{lk}(\varepsilon_{a}-\omega)|a\rangle}
{\omega-\varepsilon_{n}(1-i0)}\,.
\end{eqnarray}
The sum of the contributions (\ref{coul})-(\ref{trans2}) is 
\begin{eqnarray} \label{reccom}
\Delta E&=&\frac{1}{M}\frac{i}{2\pi}\int_{-\infty}^{\infty}d\omega\,
\langle a|(p_{i} +4\pi\alpha Z\alpha_{l} D_{li}(\omega))\nonumber\\
&&\times G(\omega +\varepsilon_{a})
( p_{i} +4\pi\alpha Z\alpha_{m} D_{mi}(\omega))|a\rangle\,.
\end{eqnarray}
This exactly coincides with  formula (\ref{yel}).

Consider now a high-$Z$ two-electron atom. For simplicity,
we will assume  that the unperturbed
wave function is a one-determinant function
\be
u(\bfx_1,\bfx_2)=\frac{1}{\sqrt{2}}\sum_{P}(-1)^P
\psi_{Pa}(\bfx_1)\psi_{Pb}(\bfx_2)\,.
\ee
The nuclear recoil correction is the sum of the one-electron
and two-electron contributions. The one-electron contribution 
is the sum of the expressions ({\ref{reccom}) for the $a$ and $b$ states.
 The two-electron
contributions are defined by the diagrams shown in Figs. 4-6.
A simple calculation of these diagrams yields 
\begin{eqnarray} \label{recint}
\Delta E^{\rm (int)}&=&
\frac{1}{M} \sum_{P}(-1)^{P}
\langle Pa|p_{i}+4\pi\alpha Z\alpha_{l}D_{li}
(\varepsilon_{Pa}-\varepsilon_{a})
|a\rangle\nonumber\\
&&\times\langle Pb|p_{i}+4\pi\alpha Z\alpha_{m}D_{mi}
(\varepsilon_{Pb}-\varepsilon_{b})
|b\rangle\,.
\end{eqnarray}
The formula (\ref{recint}) was
first derived by the quasipotential method in \cite{shabaev88}.

\section {Numerical results}

\subsection{Hydrogenlike atoms}

According to equations (\ref{recsh})-(\ref{recho})
the recoil correction is the sum of the low-order and
higher-order terms. The low-order term $\Delta E_{\rm L}$ 
is given by equation (\ref{shlo}). The higher order term 
$\Delta E_{\rm H}$ was calculated to all orders in $\alpha Z$
in \cite{artemyev95,artemyev95a,shabaev98a}. The results
of these calculations expressed in terms
of the function $P(\alpha Z)$ defined as
\be
\Delta E_{\rm H}=\frac{m^2}{M}\frac{(\alpha Z)^5}{\pi n^3}
P(\alpha Z)\,
\ee 
are presented in Table 1.
To the lowest order in $\alpha Z$ the function $P(\alpha Z)$
is given by Salpeter's expressions:
\begin{eqnarray}\label{sal1}
P_{\rm S}^{(1s)}
(\alpha Z)&=&-\frac{2}{3}\log{(\alpha Z)}-\frac{8}{3}\,
2.984129+\frac{14}{3}\log{2}
+\frac{62}{9}\,,\\
P_{\rm S}^{(2s)}(\alpha Z)&=&
-\frac{2}{3}\log{(\alpha Z)}-\frac{8}{3}\,2.811769
+\frac{187}{18}\,,\\ \label{sal2}
P_{\rm S}^{(2p_{\frac{1}{2}})}=
P_{\rm S}^{(2p_{\frac{3}{2}})}&=&
\frac{8}{3}\,0.030017-\frac{7}{18}\,.
\label{sal3}
\end{eqnarray}
Comparing the function $P(\alpha Z)$ from Table 1
 with the lowest order contributions (\ref{sal1})-(\ref{sal3})
 shows that for high Z the complete
$\alpha Z$-dependence results differ considerably from
Salpeter's ones. 

In the case of hydrogen, the difference $\Delta P=P-P_{\rm S}$
amounts to -0.01616(3), -0.01617(5), and 0.00772 for
the $1s$, $2s$, and $2p_{1/2}$ states, respectively. 
Table 2 displays the relativistic recoil corrections,
beyond the Salpeter ones, to the hydrogen energy levels.
These values  include also the corresponding
correction  from the low-order term (\ref{shlo})
which is calculated by
\begin{eqnarray}
\Delta^{\prime}E_{\rm L}^{(1s)}&=&0\,,\\
\Delta^{\prime}E_{\rm L}^{(2s)}=
\Delta^{\prime}E_{\rm L}^{(2p_{1/2})}&=&\frac{(\alpha Z)^6}{64}\;
\frac{2\,[3+\sqrt{1-(\alpha Z)^2}\,]}{[1+\sqrt{1-(\alpha Z)^2}\,]^3}
\;\frac{m^2}{M}\,.
\end{eqnarray}
The results of Refs. \cite{artemyev95,shabaev98a} which
are exact in $\alpha Z$  are compared
with the related corrections obtained to the lowest order
in $\alpha Z$. In \cite{khriplovich93,fell93} it was found
that the  $(\alpha Z)^{6}\log {(\alpha Z)}m^2/M$ corrections
cancel each other. The $(\alpha Z)^{6}m^2/M$ correction
was derived in \cite{pachucki95} for $s$-states
and in \cite{golosov95} for $p$-states. The 
 $(\alpha Z)^{7}{\rm \log}^2 (\alpha Z)m^2/M$ correction
was recently evaluated
in Refs. \cite{pachucki99,melnikov99}.
The uncertainty of the calculation based on the expansion
in $\alpha Z$ is defined by uncalculated terms of order 
$(\alpha Z)^7 m^2/M$ and is expected to be about 1 kHz for
the $1s$ state. It follows that the results of the complete
$\alpha Z$-dependence calculations are in a good agreement 
with the results obtained to lowest orders in $\alpha Z$
but  are of much higher accuracy.

As it follows from Ref. \cite{shabaev98}, the formulas (\ref{recsh})-
(\ref{recho})
will incorporate partially the nuclear size corrections
to the recoil effect if  $V_{\rm C}(r)$
 is taken to be the potential of an
extended nucleus. In particular, this replacement allows one to account
for the nuclear size corrections to the Coulomb part of 
the recoil effect.
In Ref. \cite{shabaev98b}, where the calculations of
the recoil effect for extended nuclei were performed, 
it was found that, in the case of hydrogen,
 the leading relativistic 
nuclear size correction to the Coulomb low-order part 
is comparable with the total
value of the $(\alpha Z)^{6}m^2/M$ correction
but is cancelled by the nuclear size correction
to  the Coulomb higher-order part.

One of the main goals of the calculations of Refs. 
\cite{artemyev95,artemyev95a,shabaev98b}
was to evaluate the nuclear recoil correction for highly charged ions.
In the case of the ground state of hydrogenlike uranium 
these calculations yield -0.51 eV for the point nucleus case 
\cite{artemyev95}
and -0.46 eV for the extended nucleus case \cite{shabaev98b}. 
This correction is big
enough to be included in the current theoretical prediction for
the 1s Lamb shift in hydrogenlike uranium \cite{shabaev00}
but is small compared with
the present experimental uncertainty 
which  amounts to 13 eV  \cite{stoehlker00}.
However, a much higher precision was obtained in experiments with
heavy lithiumlike ions \cite{schweppe91,beiersdorfer98}.
In this connection in Refs. \cite{artemyev95,artemyev95a} 
the nuclear recoil corrections for lithiumlike ions were calculated
as well. 

\subsection{Lithiumlike ions}

In lithiumlike ions, in addition to the one-electron contributions,
we must evaluate the two-electron contributions.
In the case of one electron over the $(1s)^2$ shell the
total two-electron
contribution  to the zeroth order in $1/Z$
is given by the expression
\be
\Delta E^{\rm int}=-\frac{1}{M}\sum_{\veps_n=\veps_{1s}}
\la a|\bfp-{\bf D}(\veps_a-\veps_n)|n\ra
\la n|\bfp-{\bf D}(\veps_a-\veps_n)|a\ra\,,
\ee
where ${\bf D}$ is defined by equation (\ref{ddd}).
Calculation of this term causes no problem 
\cite{artemyev95,artemyev95a}.
For the $2p_{1/2}$ and $2p_{3/2}$ states,
the results of this 
calculation expressed in terms of the function $Q(\alpha Z)$
defined by
\be
\Delta E^{\rm int}=-\frac{2^9}{3^8}\frac{m^2}{M}(\alpha Z)^2 
Q(\alpha Z)
\ee
are presented in Table 3.
For the $s$-states the two-electron contribution is equal zero. 
To the lowest orders in $\alpha Z$ the function $Q(\alpha Z)$
is given by \cite{shabaev94}
\be \label{qqq1}
Q_{\rm L}^{(2p_{1/2})}
(\alpha Z)&=&1+(\alpha Z)^2\Bigl(-\frac{29}{48}+
{\rm log}\frac{9}{8}\Bigr)\,,\\
Q_{\rm L}^{(2p_{3/2})}
(\alpha Z)&=&1+(\alpha Z)^2\Bigl(-\frac{13}{48}+
\frac{1}{2}{\rm log}\frac{27}{32}\Bigr)\,.
\label{qqq2}
\ee
The expressions (\ref{qqq1})-(\ref{qqq2})
serve as a good approximation for the $Q(\alpha Z)$ function
even for very high $Z$.

For low  $Z$, in addition to the corrections considered here,
the Coulomb interelectronic interaction effect on  the
non-relativistic nuclear recoil correction must be taken into
account. It contributes on the level of order 
$(1/Z)(\alpha Z)^2 m^2/M$.

To date, the highest precision in experiments with heavy ions
was obtained for the $2p_{3/2}-2s$ transition in lithiumlike
bismuth \cite{beiersdorfer98}. The transition energy measured
in this experiment amounts to $(2788.14 \pm 0.04)$ eV.
In \cite{schweppe91} the energy of the $ 2p_{1/2}-2s$ transition
in lithiumlike uranium was measured to be $(280.59 \pm 0.10)$ eV.
In both cases the recoil correction amounts to -0.07 eV and,
therefore,
is comparable with the experimental uncertainty. At present,
the uncertainty of the theoretical predictions for these 
transition energies is defined by uncalculated contributions of
second order in $\alpha$ (see Refs. \cite{shabaev00,yerokhin00}).
When calculations of these contributions are completed,
it will be possible to probe the recoil effect in high-Z
few-electron systems. This will provide a unique possibility
for testing the quantum electrodynamics in the region of
strong coupling ($\alpha Z \sim 1$) beyond the external
field approximation since in calculations of all other
 QED corrections in heavy ions
the nucleus is considered only as a stationary source
of the classical electromagnetic  field. 

%%%%%%%%%%%%%%%%%%%%%%%%    
\section{Conclusion}    
%%%%%%%%%%%%%%%%%%%%%%%%   
In this paper the relativistic theory of the recoil effect in atoms
is considered. It is shown that the complete $\alpha Z$-dependence
calculation of the recoil correction provides the highest precision 
even in the case of hydrogen. The recoil
corrections to the energy levels of highly charged ions
contribute on the level of the present experimental accuracy.
It provides good perspectives for testing the quantum electrodynamics
in  the region of strong coupling ($\alpha Z \sim 1$) beyond
the external field approximation.

\section*{Acknowledgments}    
The author wants to express his thanks to    
 A.N. Artemyev, T. Beier, G. Plunien, G. Soff, and V.A. Yerokhin 
for stimulating  collaboration. 
Valuable conversations with S.G. Karshenboim, P.J. Mohr,
 and A.S. Yelkhovsky   
 are gratefully  acknowledged.    

%%%%%%%%%%%%%%%%%%%%%%%%     
    
%%%%%%%%%%%%%%%%%%%%%%%%%%%%%%   

\newpage

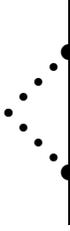
\begin{figure}
\setlength{\unitlength}{1mm}
\begin{center}
\begin{picture}(100,60)(-5,0)
   \put(40,10){\line(0,1){30}}
   \multiput(40,17)(-2,2){5}{\circle*{1}}
   \multiput(40,33)(-2,-2){5}{\circle*{1}}
   \put(40,33){\circle*{2}}
   \put(40,17){\circle*{2}}
   \end{picture}
   \end{center}
   \begin{center}
\caption{ Coulomb nuclear recoil diagram.}
\end{center}
\end{figure}

\begin{figure}
\setlength{\unitlength}{1mm}
\begin{center}
\begin{picture}(100,60)(-5,0)
   \put(20,10){\line(0,1){30}}
  \put(60,10){\line(0,1){30}}
   \multiput(20,17)(-2,2){5}{\line(-1,0){1}}
   \multiput(20,33)(-2,-2){5}{\line(-1,0){1}}
   \multiput(60,17)(2,2){5}{\line(1,0){1}}
   \multiput(60,33)(2,-2){5}{\line(1,0){1}}
   \put(60,33){\circle*{2}}
   \put(20,17){\circle*{2}}
   \put(18,1){a}
   \put(58,1){b} 
   \end{picture}
   \end{center}
   \begin{center}
\caption{ One-transverse-photon nuclear recoil diagrams.}
\end{center}
\end{figure}

\begin{figure}
\setlength{\unitlength}{1mm}
\begin{center}
\begin{picture}(100,60)(-5,0)
   \put(40,10){\line(0,1){30}}
   \multiput(40,17)(-2,2){4}{\line(-1,0){1}}
   \multiput(40,33)(-2,-2){4}{\line(-1,0){1}}
   \put(33,25){\circle*{2}}
   \end{picture}
   \end{center}
\begin{center}
\caption{ Two-transverse-photon nuclear recoil diagram.}
\end{center}
\end{figure}

\begin{figure}
\setlength{\unitlength}{1mm}
\begin{center}
\begin{picture}(100,60)(-15,0)
  \put(20,10){\line(0,1){30}}
   \put(40,10){\line(0,1){30}}
   \multiput(20,25)(2,0){10}{\circle*{1}}
   \put(20,25){\circle*{2}}
   \put(40,25){\circle*{2}}
   \end{picture}
\end{center}
\begin{center}
\caption{Two-electron Coulomb nuclear recoil diagram.}
\end{center}
\end{figure}
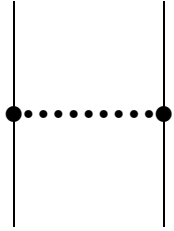
\begin{figure}
\setlength{\unitlength}{1mm}
\begin{center}
\begin{picture}(120,60)(-5,0)
  \put(20,10){\line(0,1){30}}
  \put(40,10){\line(0,1){30}}
  \multiput(20,25)(2,0){10}{\line(1,0){1}}
   \put(20,25){\circle*{2}}
   \put(27,2){a}
   \put(67,2){b} 
  \put(60,10){\line(0,1){30}}
  \put(80,10){\line(0,1){30}}
  \multiput(60,25)(2,0){10}{\line(1,0){1}}
   \put(80,25){\circle*{2}}
\end{picture}
\end{center}
\begin{center}
\caption{Two-electron one-transverse-photon nuclear recoil diagrams.}
\end{center}
\end{figure}
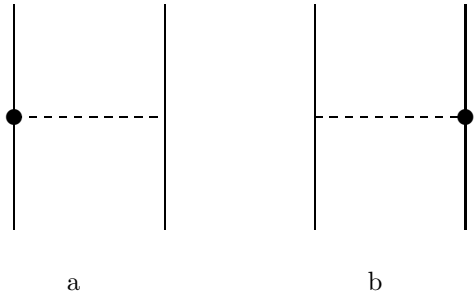
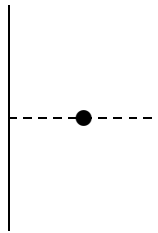
\begin{figure}
\setlength{\unitlength}{1mm}
\begin{center}
\begin{picture}(100,60)(-15,0)
  \put(20,10){\line(0,1){30}}
   \put(40,10){\line(0,1){30}}
   \multiput(20,25)(2,0){10}{\line(1,0){1}}
   \put(30,25){\circle*{2}}
   \end{picture}
   \end{center}
   \begin{center}
\caption{Two-electron two-transverse-photon nuclear recoil diagram.}
\end{center}
\end{figure}

\newpage

\begin{table}
\caption{The  results of the numerical calculation of the
function $P(\alpha Z)$ for low-lying states of hydrogenlike atoms.
}
\vspace{0.5cm}
\begin{center}
\begin{tabular}{c l l l l l }  \hline
$\;\;\;\;Z\;\;\;\;$&$1s\;\;\;\;\;\;\;\;\;\;\;\;\;\;\;\;\;\;\;
$&$2s\;\;\;\;\;\;\;\;\;\;\;\;\;\;\;\;\;\;\;\;$
&$2p_{1/2}\;\;\;\;\;\;\;\;\;\;\;\;\;\;\;\;\;\;$&$2p_{3/2}$\\ \hline
1&5.42990(3)&6.15483(5)&-0.30112&-0.3013(4)$\;\;$\\ 
5&4.3033(4)&5.0335(2)&-0.2692&-0.2724(1)\\ 
10&3.7950(1)&4.5383(1)&-0.2277&-0.2379\\ 
20&3.2940(1)&4.0825&-0.1393&-0.1726\\ 
30&3.0437(1)&3.9037&-0.0421&-0.1107\\ 
40&2.9268(1)&3.8900&0.0685&-0.0517\\ 
50&2.9137(1)&4.0228(1)&0.2000&0.0050\\ 
60&3.0061(2)&4.3248(2)&0.3655&0.0597\\ 
70&3.2334(4)&4.8656(5)&0.5894&0.1125\\ 
80&3.672(1)&5.807(2)&0.9214(2)&0.1638\\ 
90&4.519(8)&7.557(9)&1.481(1)&0.2138\\ 
100&6.4(1)&11.4(2)&2.63(2)&0.2625\\ \hline 
\end{tabular}
\end{center}
\end{table}

\begin{table}
\caption{The values of the relativistic recoil correction
to hydrogen energy levels beyond the Salpeter contribution, in kHz.
The values given in the second and third rows
include the $(\alpha Z)^6m^2/M$ contribution and all the
 contributions of higher orders in $\alpha Z$. In the last row
the sum of the $(\alpha Z)^6m^2/M$ and $(\alpha Z)^7
{\rm log}^2(\alpha Z)m^2/M$ contributions is  given.}
\vspace{0.5cm}
\begin{center}
\begin{tabular}{l l l l}  \hline
$\;\;\;$State $\;\;\;\;\;\;\;\;\;\;\;\;\;\;\;\;\;\;\;\;\;\;\;\;\;\;\;
\;\;\;\;\;\;\;\;\;\;\;\;\;\;\;\;\;\;\;\;\;\;\;\;\;\;\;\;\;\;\;\;\;\;
$
& $1s\;\;\;\;\;\;\;\;\;\;\;\;\;\;\;\;\;\;\;$ & $2s\;\;\;\;\;\;\;\;\;
\;\;\;\;\;\;\;\;\;\;$ &$2p_{1/2}\;\;\;\;$\\ \hline
$\;\;\;$To all orders in $\alpha Z$, Ref. [25]  
&-7.1(9) &-0.73(6) &0.59 \\ 
$\;\;\;$To all orders in $\alpha Z$, Ref. [27] 
&-7.16(1)  & -0.737(3)  & 0.587 \\ 
$\;\;\;(\alpha Z)^6m^2/M$, Refs. [18,30] &-7.4  & -0.77  & 0.58 \\ 
$\;\;\;(\alpha Z)^7 {\rm log}^2(\alpha Z)m^2/M$, Refs. [31,32] 
& -0.4  & -0.05  &  \\ 
$\;\;\;$The sum of the low-order terms & -7.8  & -0.82  &  \\ \hline
\end{tabular}
\end{center}
\end{table} 

\begin{table}
\caption{The  results of the numerical calculation of the
function $Q(\alpha Z)$ for low-lying states of lithiumlike ions.}
\vspace{0.5cm}
\begin{center}
\begin{tabular}{l l l}  \hline
$\;\;\;\;Z\;\;\;\;\;\;\;\;\;\;\;\;\;\;\;\;\;\;
$&$(1s)^2 2p_{1/2}\;\;\;\;\;\;\;\;\;\;\;\;\;\;\;$
&$(1s)^2 2p_{3/2}\;\;$\\ \hline
$\;\;\;\;$10&0.99741&0.99810\\ 
$\;\;\;\;$20&0.98959&0.99239\\ 
$\;\;\;\;$30&0.97645&0.98281\\ 
$\;\;\;\;$40&0.95776&0.96926\\ 
$\;\;\;\;$50&0.93313&0.95165\\ 
$\;\;\;\;$60&0.90195&0.92988\\ 
$\;\;\;\;$70&0.86320&0.90390\\ 
$\;\;\;\;$80&0.81529&0.87362\\ 
$\;\;\;\;$90&0.75570&0.83896\\ 
$\;\;\;$100&0.68041&0.79951\\ \hline
\end{tabular}
\end{center}
\end{table}


\newpage     
\begin{thebibliography}{99}    
\addcontentsline{toc}{section}{References}
\bibitem{breit48}     
G. Breit, G.E. Brown: Phys. Rev. {\bf 74},    
1278 (1948)     
\bibitem{bechert35}    
K. Bechert, J. Meixner: Ann. Phys., Lpz.  {\bf 22},    
525 (1935)    
\bibitem{salpeter51}
E.E. Salpeter and H.A. Bethe: Phys. Rev. {\bf 84}, 1232 (1951)
\bibitem{logunov63}
A.A. Logunov and A.N. Tavkhelidze:
 Nuovo Cimento {\bf 29}, 380 (1963)
\bibitem{salpeter52}
E.E. Salpeter: Phys. Rev. {\bf 87}, 328 (1952)
\bibitem{bethe57}
H.A. Bethe,  E.E. Salpeter: {\it Quantum Mechanics of One- and Two-Electron
Atoms} (Springer, Berlin, 1957)
\bibitem{hansch00}
F. Biraben and T. W. H\"ansch: this volume
\bibitem{schweppe91}
J. Schweppe, A. Belkacem, L. Blumenfeld, N. Claytor, B. Feinberg,
H. Gould, V.E. Kostroun, L. Levy, S. Misawa, J.R. Mowat, 
M.H. Prior: Phys. Rev. Lett. {\bf 66}, 1434 (1991)
\bibitem{beiersdorfer98}    
P. Beiersdorfer, A. Osterheld, J. Scofield, J.R. Crespo Lopez-Urrutia,    
 K. Widmann: Phys. Rev. Lett. {\bf 80}, 3022 (1998)    
\bibitem{stoehlker00}
T. St\"ohlker: this volume
\bibitem{labzowsky72}
L.N. Labzowsky: In: Papers at 17th All-Union Symposium on Spectroscopy
(Astrosovet, Moscow, 1972), Part 2, pp. 89-93
\bibitem{braun73}
M.A. Braun: Sov. Phys. JETP {\bf 37}, 211 (1973)
\bibitem{shabaev98}
V.M. Shabaev: Phys. Rev. A {\bf 57}, 59 (1998)
\bibitem{shabaev85}
V.M. Shabaev: Theor. Math. Phys. {\bf 63}, 588 (1985);
In: Papers at First Soviet-British Symposium on Spectroscopy
of Multicharged Ions (Academy of Sciences, Troitsk, 1986), pp. 238-240
\bibitem{gross69}
F. Gross: Phys. Rev. {\bf 186}, 1448 (1969)
\bibitem{faustov75}
L.S. Dul'yan, R.N. Faustov: Teor. Mat. Fiz. {\bf 22}, 314 (1975)
\bibitem{yelkhovsky94}
A.S. Yelkhovsky: Preprint BINP 94-27 
(Budker Inst. of Nuclear Physics, Novosibirsk, 1994);
hep-th/9403095 (1994); JETP {\bf 83}, 230 (1996)
\bibitem{pachucki95}
K. Pachucki, H. Grotch: Phys. Rev. A {\bf 51}, 1854 (1995)
\bibitem{epstein62}
J. Epstein, S. Epstein: Am. J. Phys. {\bf 30}, 266 (1962)
\bibitem{shabaev91}
V.M. Shabaev: J. Phys. B {\bf 24}, 4479 (1991)
\bibitem{shabaev88}
V.M. Shabaev: Sov. J. Nucl. Phys. {\bf 47} 69 (1988)
\bibitem{palmer87} 
C.W. Palmer: J. Phys. B {\bf 20}, 5987 (1987)
\bibitem{shabaev94}
V.M. Shabaev, A.N.Artemyev: J. Phys. B {\bf 27}, 1307 (1994)
\bibitem{shabaev94a}
V.M. Shabaev, I.G. Fokeeva: Phys. Rev. A {\bf 49}, 4489 (1994);
V.M. Shabaev: Phys. Rev. A {\bf 50}, 4521 (1994)
\bibitem{artemyev95}
A.N. Artemyev, V.M. Shabaev, V.A. Yerokhin:
Phys. Rev. A {\bf 52}, 1884 (1995)
\bibitem{artemyev95a}
A.N. Artemyev, V.M. Shabaev, V.A. Yerokhin:
 J.Phys. B {\bf 28}, 5201 (1995)
\bibitem{shabaev98a}
V.M. Shabaev, A.N. Artemyev, T. Beier, G. Soff:
J. Phys. B {\bf 31}, L337 (1998)
\bibitem{khriplovich93}
I.B. Khriplovich, A.I. Milstein, A.S. Yelkhovsky:
 Phys. Scr. T {\bf 46}, 252 (1993)
\bibitem{fell93}
R.N. Fell, I.B. Khriplovich, A.I. Milstein,
A.S. Yelkhovsky: Phys. Lett. A {\bf 181},
172 (1993)
\bibitem{golosov95}
E.A. Golosov, I.B. Khriplovich, A.I. Milstein, A.S. Yelkhovsky:
JETP {\bf 80}, 208 (1995)
\bibitem{pachucki99}
K. Pachucki, S. Karshenboim: Phys. Rev. A {\bf 60}, 2792 (1999)
\bibitem{melnikov99}
K. Melnikov, A. Yelkhovsky: Phys. Lett. B {\bf 458}, 143 (1999)
\bibitem{shabaev98b}
V.M. Shabaev, A.N. Artemyev, T. Beier, G. Plunien, V.A. Yerokhin,
G. Soff: Phys. Rev. A {\bf 57}, 4235 (1998);
 Phys. Scr. T {\bf 80}, 493 (1999)
\bibitem{shabaev00}
V.M. Shabaev, A.N. Artemyev, V.A. Yerokhin: Phys. Scr. T {\bf 86},
7 (2000)
\bibitem{yerokhin00}
V.A. Yerokhin, A.N. Artemyev, V.M. Shabaev, M.M. Sysak, 
O.M. Zherebtsov, G. Soff: to be published
    
\end{thebibliography}
\end{document}